\documentclass{vgtc}                          %

\ifpdf%
  \pdfoutput=1\relax                   %
  \usepackage{graphicx}                %
  \DeclareGraphicsExtensions{.pdf,.png,.jpg,.jpeg} %
\else%
  \usepackage{graphicx}                %
  \DeclareGraphicsExtensions{.eps}     %
\fi%

\graphicspath{{figures/}{pictures/}{images/}{./}} %

\usepackage{microtype}                 %
\PassOptionsToPackage{warn}{textcomp}  %
\usepackage{textcomp}                  %
\usepackage{mathptmx}                  %
\usepackage{times}                     %
\usepackage{cite}                      %
\usepackage{tabu}                      %
\usepackage{booktabs}                  %

\onlineid{0}

\vgtccategory{Research}

\vgtcinsertpkg

\title{Development of an Immersive Virtual Colonoscopy Viewer\\ for Colon Growths Diagnosis}

\author{João Serras\thanks{e-mail: nuno591lopes@gmail.com}\\ %
        \scriptsize IST - ULisboa
\and Anderson Maciel\thanks{e-mail: anderson.maciel@tecnico.ulisboa.pt}\\ %
     \scriptsize IST - ULisboa, INESC-ID
\and Soraia Paulo\thanks{e-mail: soraiafpaulo@tecnico.ulisboa.pt}\\ %
     \scriptsize IST - ULisboa, INESC-ID
\and Andrew Duchowski\thanks{e-mail: aduchow@clemson.edu}\\ %
    \scriptsize Clemson University%
\vspace{6pt}
\and Regis Kopper\thanks{e-mail: kopper@uncg.edu}\\ %
 \scriptsize UNC Greensboro %
\and Catarina Moreira\thanks{e-mail: catarina.pintomoreira@qut.edu.au}\\ %
 \scriptsize Queensland University of Technology %
\and Joaquim Jorge\thanks{e-mail: jorgej@tecnico.ulisboa.pt}\\ %
\scriptsize IST - ULisboa, INESC-ID %
}

\teaser{
  \centering
  \includegraphics[width = 200pt]{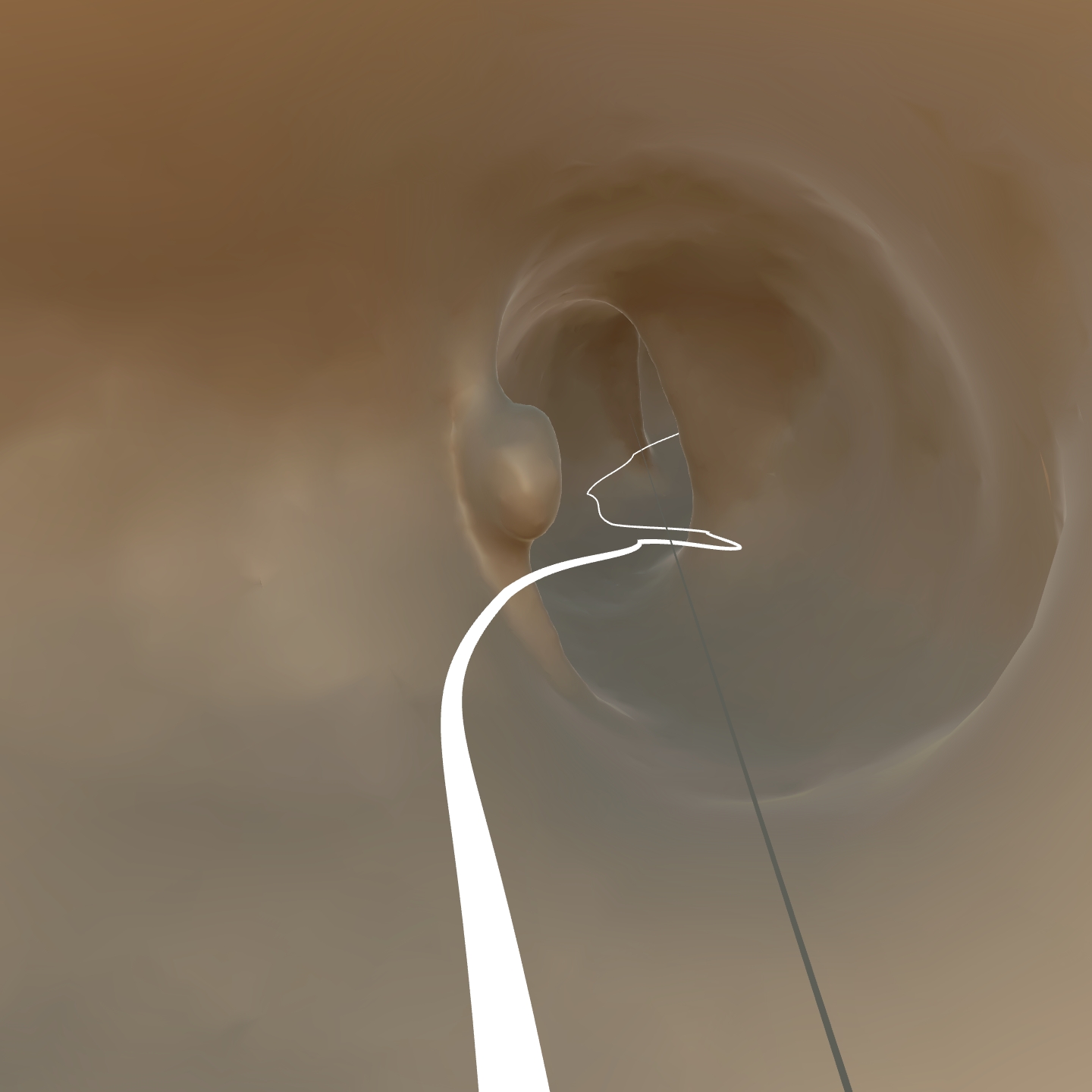}
  \includegraphics[height=200pt]{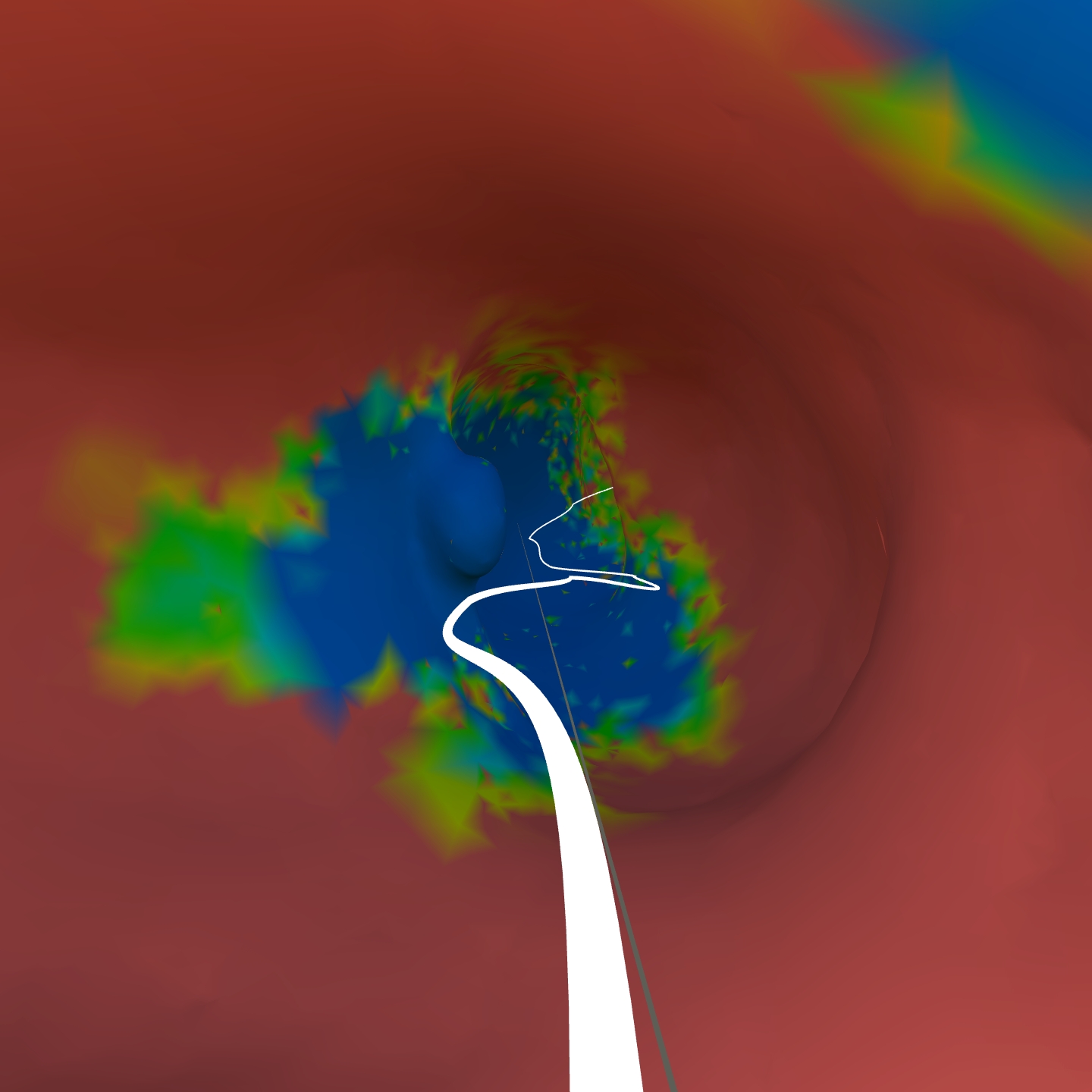}
  \caption{Left: view from inside the colon in VR. Right: a colormap view indicating the areas already examined in blue and not yet seen in red.}
  \label{fig:teaser}
}

\abstract{Desktop-based virtual colonoscopy has been proven to be an asset in the identification of colon anomalies. The process is accurate, although time-consuming. The use of immersive interfaces for virtual colonoscopy is incipient and not yet understood. In this work, we present a new design exploring elements of the VR paradigm to make the immersive analysis more efficient while still effective. We also plan the conduction of experiments with experts to assess the multi-factor influences of coverage, duration, and diagnostic accuracy.%
} %

\CCScatlist{
  \CCScatTwelve{Immersive Virtual Colonoscopy}{Visu\-al\-iza\-tion}{Visu\-al\-iza\-tion techniques};
  \CCScatTwelve{Immersive Virtual Colonoscopy}{Visu\-al\-iza\-tion}{Visualization design and evaluation methods}{}
}

\CCScatlist{
  \CCScatTwelve{Human-centered computing}{Human computer interaction (HCI)}{nteraction paradigms}{Virtual reality};
  \CCScatTwelve{Applied computing}{Life and medical sciences}{Computational biology}{Imaging}
}

\begin{document}

\firstsection{Introduction}

\maketitle

\begin{figure}[ht]
\centering
\includegraphics[width = 240pt]{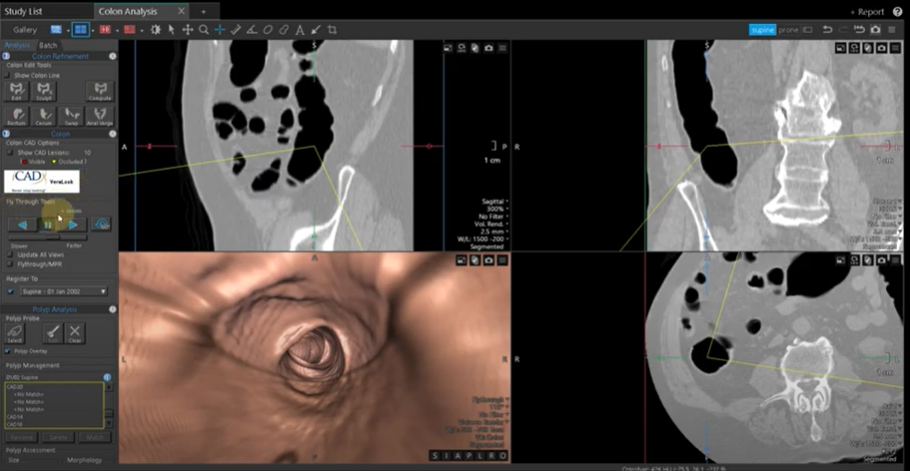}
\caption{Desktop-based virtual colonoscopy in a 2x2 view configuration.}
\label{fig_desktop}
\end{figure}

In 2020, Colorectal Cancer (CRC) was  ranked third in terms of incidence and second as the cause of cancer-related deaths \cite{ColonandRectalPolyps}.
The field of medical imaging has made significant advancements in recent years, offering new and improved ways to diagnose and treat various conditions. One such innovation is the development of CT colonography, also known as virtual colonoscopy. Unlike other CT-based exams, the diagnosis of colon polyps and diverticles can hardly be done solely on sliced image volumes. Instead, the internal surface of the colon is reconstructed in 3D and displayed on the radiologist monitor from the perspective of an observer at the colon mid-line. This, combined with a very wide field of view projection, allows for tunnel-like navigation (see fig.~\ref{fig_desktop}) that makes colon growths apparent, permitting experts to measure them. 

Virtual reality has been applied to scientific and data visualization over two decades with exciting results~\cite{progressReport,surveyIA22}. Several features of VR, such as presence, natural look around, navigation, and manipulation, among others, are responsible for a more focused and efficient visualization task when the data has multiple dimensions. 

In this context, we propose the development of an immersive virtual colonoscopy viewer for the diagnosis of colon growths. Such viewer utilizes state-of-the-art technology to create a highly realistic and interactive 3D representation of the colon, allowing doctors to identify and examine growths in greater detail than ever before. This approach may improve the accuracy and/or efficiency of colon growth diagnoses while offering a more comfortable and less invasive alternative to traditional colonoscopy procedures and current desktop implementations.

The translation of a desktop viewer for an immersive one sounds quite straightforward. Nevertheless, very few works in the literature approach diagnostic colon visualization with an immersive interface, and none provide consistent guidelines for designing this visualization and the related interactive actions.

In this paper, we present preliminary results from the development of a VR app that allows an immersed user to enter the virtual colon and systematically explore its whole length for clinically relevant findings. Our VR app also benefits from eye tracking as well as a set of VR metaphors with the goal of making the radiologist's job at least as accurate and less time-consuming to help make the procedure more widely available for patients.

\section{Background and Related Work}

Nowadays, Total Colonoscopy (TC) is the most accurate method to detect colorectal cancer. Despite being so, it entails many disadvantages and complications, such as discomfort, need for preparations, risk of rupture, and many others due to its invasiveness, since it requires a scope to be inserted through the rectum.

Virtual Colonoscopy (VC) was created to solve some of TC disadvantages. Through the use of CT scans and desktop programs, radiologists navigate through a reconstructed colon in a 2-way path: from the rectum to the cecum and from the cecum to the rectum to avoid missing anomalies in folds. Despite solving invasiveness, it brings its own disadvantages. Navigation is restricted to a single view, causing the level of uncertainty regarding the types of polyps to be higher, which in turn leads to a higher referral rate to colonoscopy and superior cost. While VC is able to detect most colon anomalies, it is highly time-consuming, taking much longer than a conventional total colonoscopy. Moreover, the exam only takes into account the inside of the colon; extracolonic lesions may still exist. Thus the need for further exams increases the cost, and on top of that, false positives can occur.

Finally, Immersive Virtual colonoscopy(IVC) was proposed. A 3D model of the colon is reconstructed from CT scans, similar to VC. However, the resulting 3D colon surface is viewed within a virtual reality environment. IVC allows radiologists to freely explore the inside and outside of the colon using natural head and body motion, better analyze volumetric data with stereopsis and parallax cues, and still have the same functionalities as VC. Since the radiologist is not bound to a single path and view, many possibilities are created. Measures can be taken directly in 3D. Despite all these possibilities, radiologists are not interested in fancy VR interaction. The major problem we face in designing a VR interface for IVC is its efficiency. Thus, the major challenge we face is to reduce the duration of the exam keeping the fidelity high. Later in the paper, we describe how we designed the interactions so that the exam can be made more efficiently.

\subsection{Anomalies}
The colon exams described have the same objective: finding and categorizing anomalies in
the colon. Anyone can develop these anomalies that are distinguished in 5 types:

\textbf{Adenomatous (tubular adenoma)} - Most common type of colon polyp and mostly benign. They can be transformed into other cancerous types although this process can take several years and can be prevented with regular screening;

\textbf{Serrated} - This type of polyp can become cancerous or not depending on its size and location. If large, typically flat and found in the upper colon, they are precancerous. This kind, due to its characteristics, is hard to locate.

\textbf{Hyperplastic} - If a serrated polyp is small and found in the lower colon, they are known as Hyperplastic; they are rarely malignant.

\textbf{Inflammatory} - Occurs mostly in people who have inflammatory bowel. Also
known as pseudopolyps. They are a product of chronic inflammation in the colon. Inflammatory polyps are benign and often do not turn into cancerous types.

\textbf{Villous Adenoma (Tubulovillous Adenoma)} - Commonly sessile (flat), this type of polyp has a high risk of becoming cancerous. Small polyps can be removed during a colonoscopy while larger ones may require additional surgery to complete their removal.

All types of Polyps should be identified but adenomas, in particular, are found to turn into cancerous polyps more often. When they are flat, they can be hard to identify. Furthermore, serrated polyps are a strong factor to CRC as the polyp may be too small and pass undetected, especially when preparations were not done properly\cite{ColonandRectalPolyps}.

With the necessities of the radiologists in mind, previous works studied how VR impacts the diagnosis~\cite{venson2018case}. Then, a few papers endeavored to improve their experience using spatially aware displays~\cite{grandi2010interactive} and immersive VR. Firstly, to help navigation in VR, previous papers have shown that World in Miniature and teleport are great ways to reduce time consumption and improve orientation while inside cave-like environments such as a colon~\cite{WIMRELATEDWORK,WIM}. Secondly, a comparison of navigation methods through a path inside the colon has shown that, despite "Fly-over" having the best accuracy, "Fly-through" was the most liked~\cite{ICTFIC}.

Regarding the interface and interactions, there are many interaction and selection techniques available and studied. In this project, we decided to use ray-based techniques~\cite{selectionTechniques,selectionInterface}. A similar system was previously developed \cite{IVC}, but at the time the HMD used did not have enough resolution and received significant cybersickness complaints. 

Another technology we rely on is eye-tracking. There are previous works regarding eye-tracking tools for medical purposes. Studies have shown that they bring significant benefits as visualization aids. They have also suggested and explained metrics in order to evaluate such tools~\cite{Eye-tracking}.

\section{IVC Design}

In VC radiologists are used to be able to navigate the colon through its center line at different speeds, bookmark suspected regions, measure regions, list areas measured, and inspect the CT scans of a suspected region to define the diagnosis. Sometimes a droplet of mucus is shaped like a polyp. Density values available in CT images are used to disambiguate.

In our IVC design, we tried to support the same tasks but with VR elements and metaphors.

\begin{figure}[ht]
\centering
\includegraphics[width = 240pt,height = 170pt]{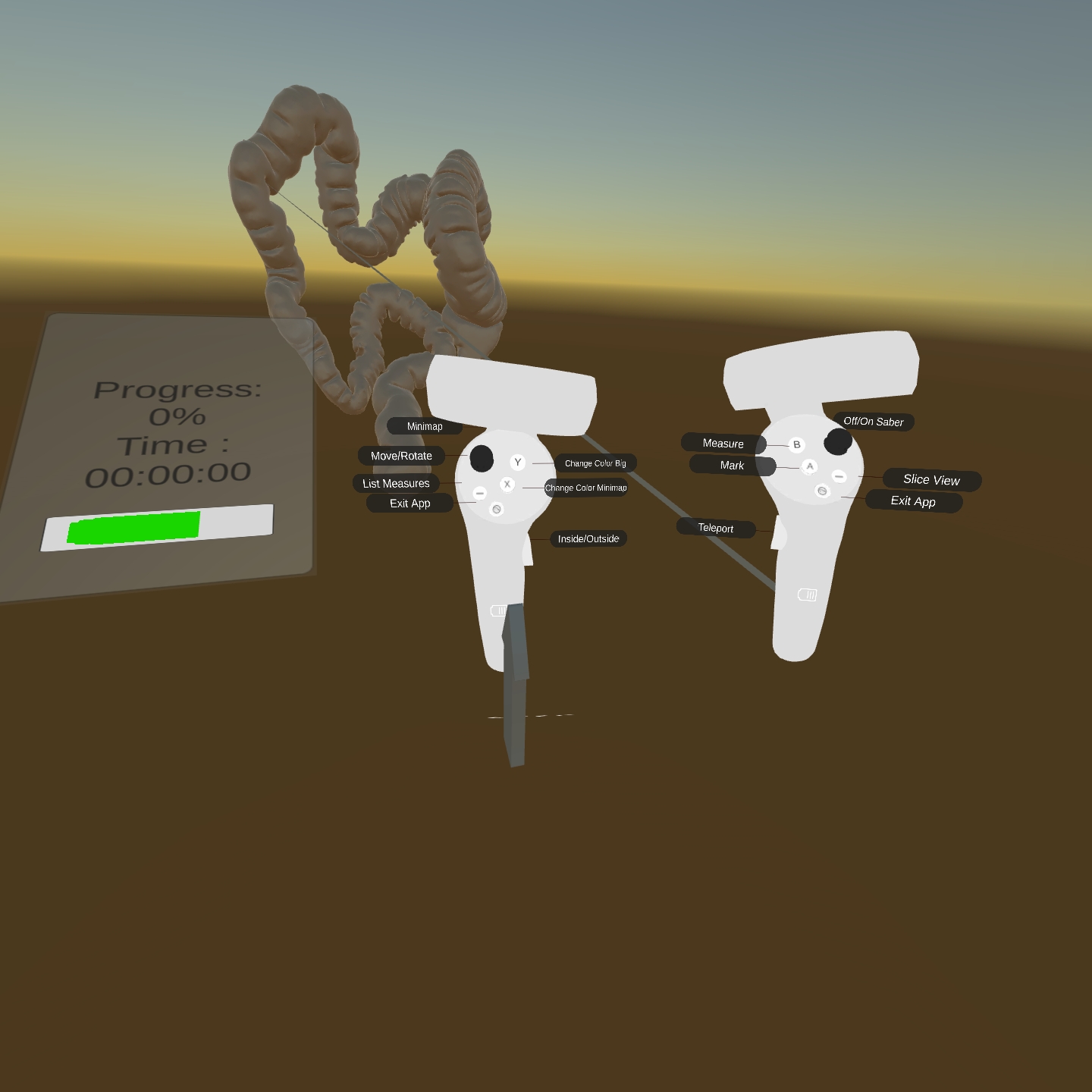}
\caption{Virtual environment outside the colon showing the controls.}
\label{fig_outside}
\end{figure}

The VE contains an infinite horizon landscape where the virtual colon is floating. Upon entrance into the VE, the user sees the colon from outside in an anatomical position. They also see the avatars of the two controllers they are handling, with labels indicating how to operate. With the left hand, the user can rotate the colon model, and with the right hand, they can point to a location on the colon where they want to start the analysis. Fig.~\ref{fig_outside} illustrates this first view.

The radiologist generally navigates from the rectum to the colon due to training with conventional techniques. So, let us suppose that they pointed toward the rectum and teleported there. From there, the radiologist navigates the colon using the fly-through technique through the mid-line. Five different velocities are chosen naturally with the joystick of the right hand. The motion direction is the tangent of the mid-line, but the radiologist can look around and, when looking back, the navigation direction is inverted. Small natural head and body movements are still allowed, which helps quickly approach the colon wall to see details or tilt the head sideways to better grasp depth features.

\begin{figure}[ht]
\centering
\includegraphics[width = 240pt,height = 170pt]{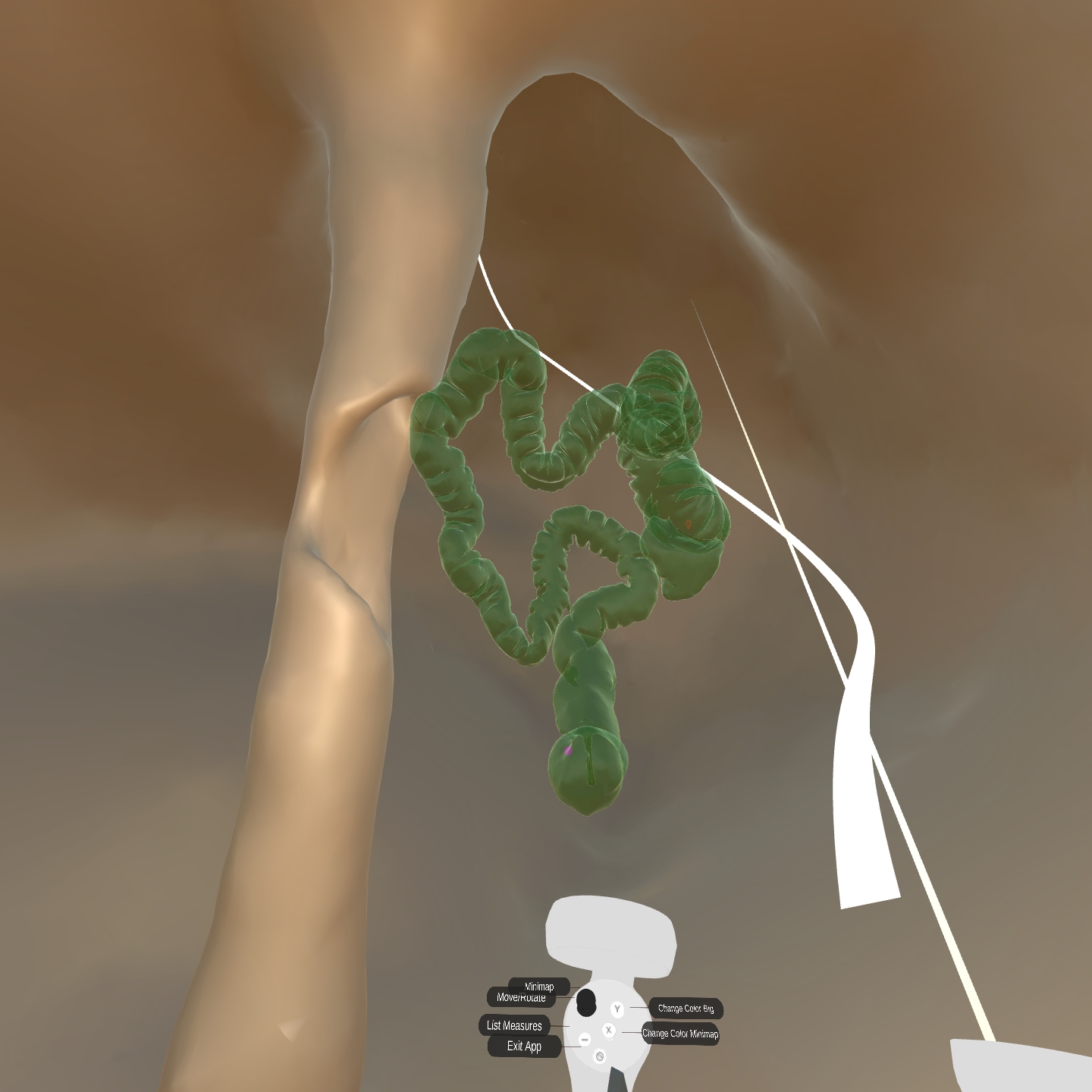}
\caption{World in miniature attached to the hand.}
\label{fig_wim}
\end{figure}

Several cues help with self-orientation. First, anatomical elements allow the specialist to recognize the portion of the colon they are seeing. Second, we provide an arrow object that always points vertically. Finally, a world in miniature (WIM) depicting an external view of the colon can be toggled and appears attached to the left hand (fig.~\ref{fig_wim}). It is possible to teleport to other locations either pointing on the WIM or pointing at the colon itself. The teleport destination is the closest position in the mid-line path to the first ray intersection. The line shown at the mid-line path is originally white but will be painted green at the segments already visited.

\begin{figure}[ht]
\centering
\includegraphics[width = 240pt,height = 170pt]{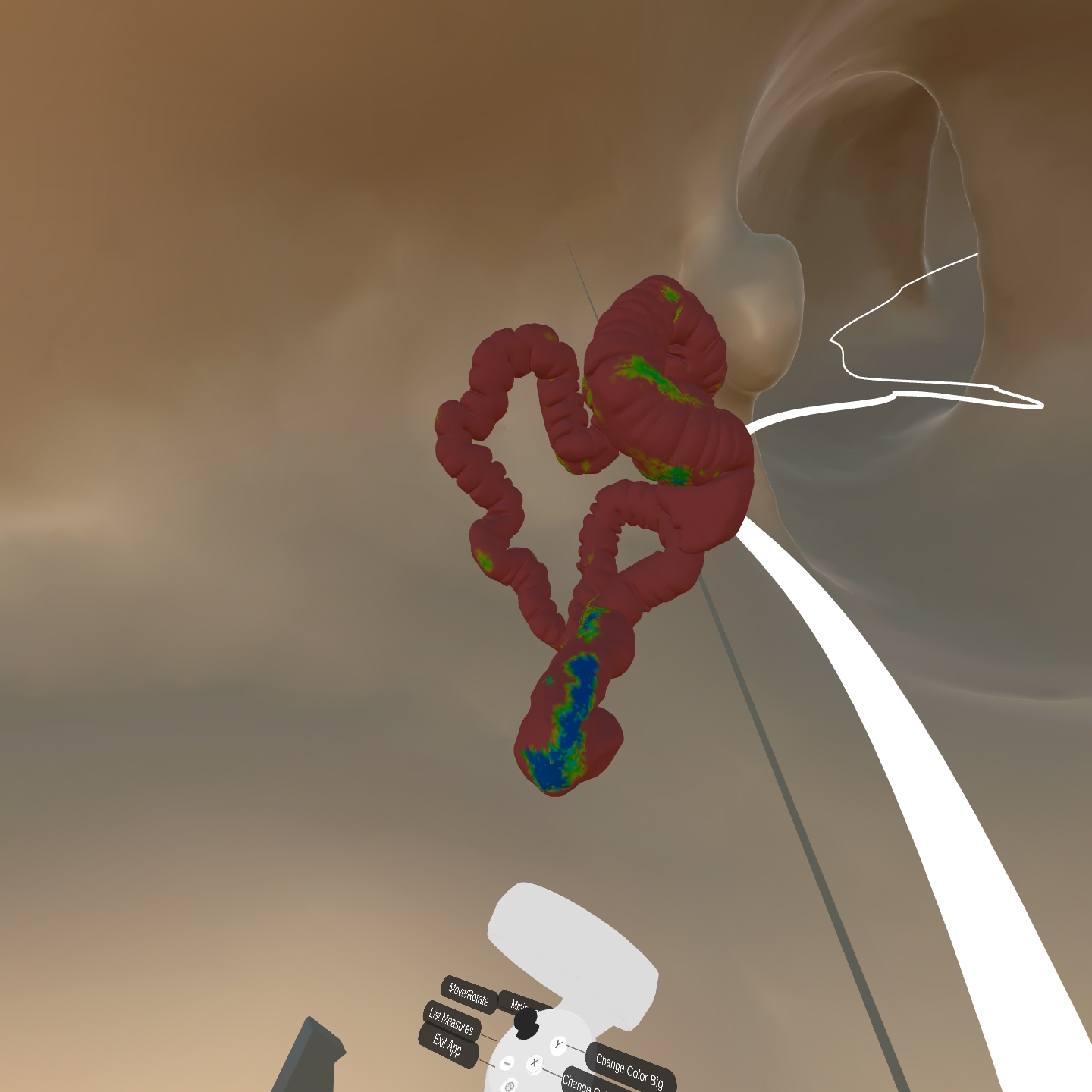}
\caption{Coverage heatmap indicating on the world in miniature for how long each area of the colon was observed.}
\label{fig_heatmapWIM}
\end{figure}

An additional feature, without parallel in the desktop app, is the ability to compute the time the user spent looking at each point on the colon wall. Using the center of the viewport as a reference, we project a set of vectors toward the model and add time to the points intercepted. When eye tracking is available, which is the case with our prototype, the viewport center approach is replaced by the respective eye point in the viewport, which increases the reliability of the times computed. Time is then mapped to colors, and a heatmap can be consulted by the radiologist if they want to make sure nothing was left out. The heatmap can be applied either for the WIM (fig.~\ref{fig_heatmapWIM}) or the colon model itself (fig.~\ref{fig:teaser}). Moreover, the view time for each point is valuable information for future validation of the VR interface and for meta-analysis of reading patterns.

\begin{figure}[ht]
\centering
\includegraphics[width = 240pt,height = 170pt]{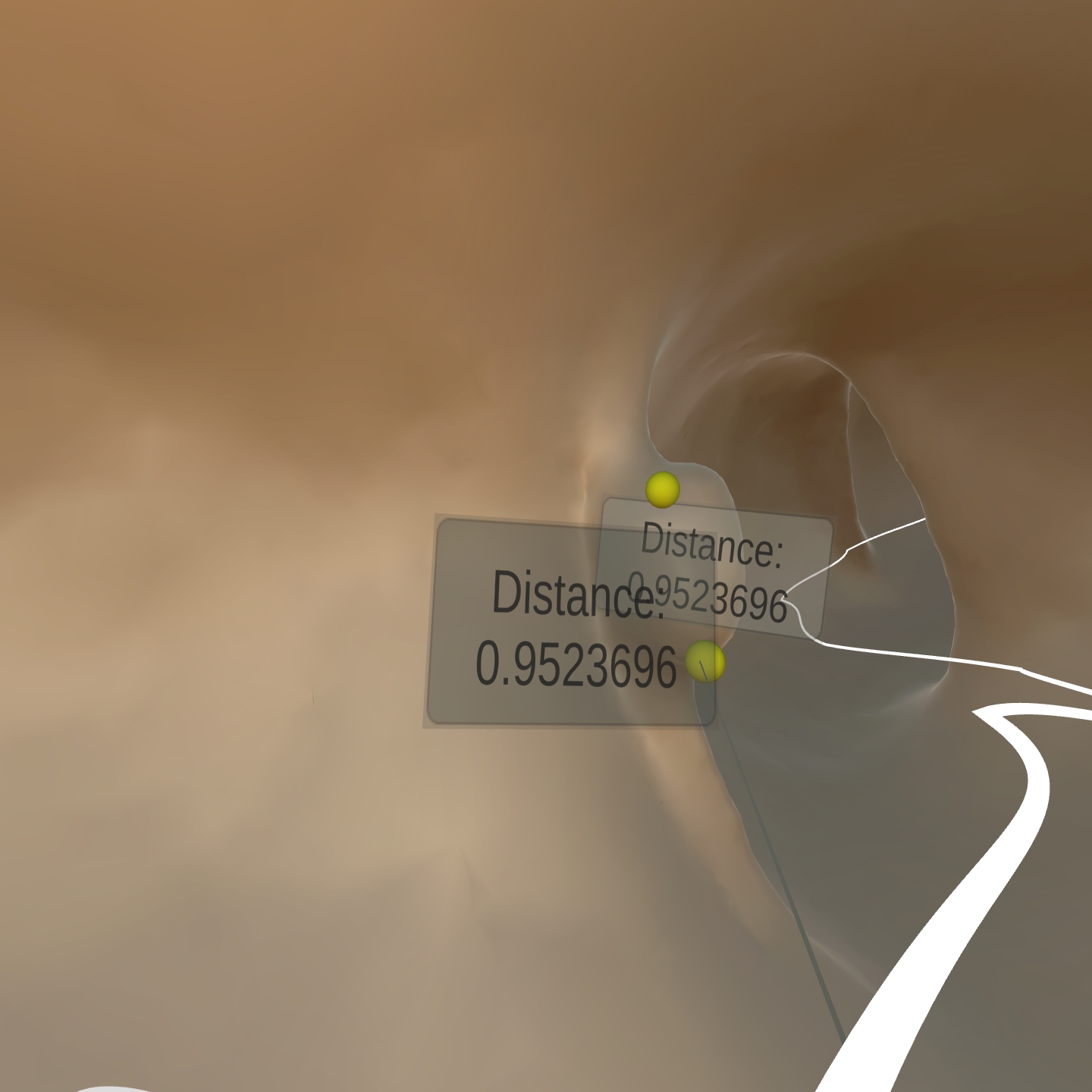}
\caption{Distance measurement tool to size the anomalies.}
\label{fig_measure}
\end{figure}

The interface allows one to point and bookmark regions for later analysis and categorization. A list of bookmarks is available where a selection teleports the user to the respective location for quick access. Besides bookmarking, measurement is possible in 3D. While in desktop apps the measures are taken on the three planes, we offer a tool to select any two points on the colon and measure the distance between them (see fig.~\ref{fig_measure}).

\begin{figure}[ht]
\centering
\includegraphics[width = 240pt,height = 170pt]{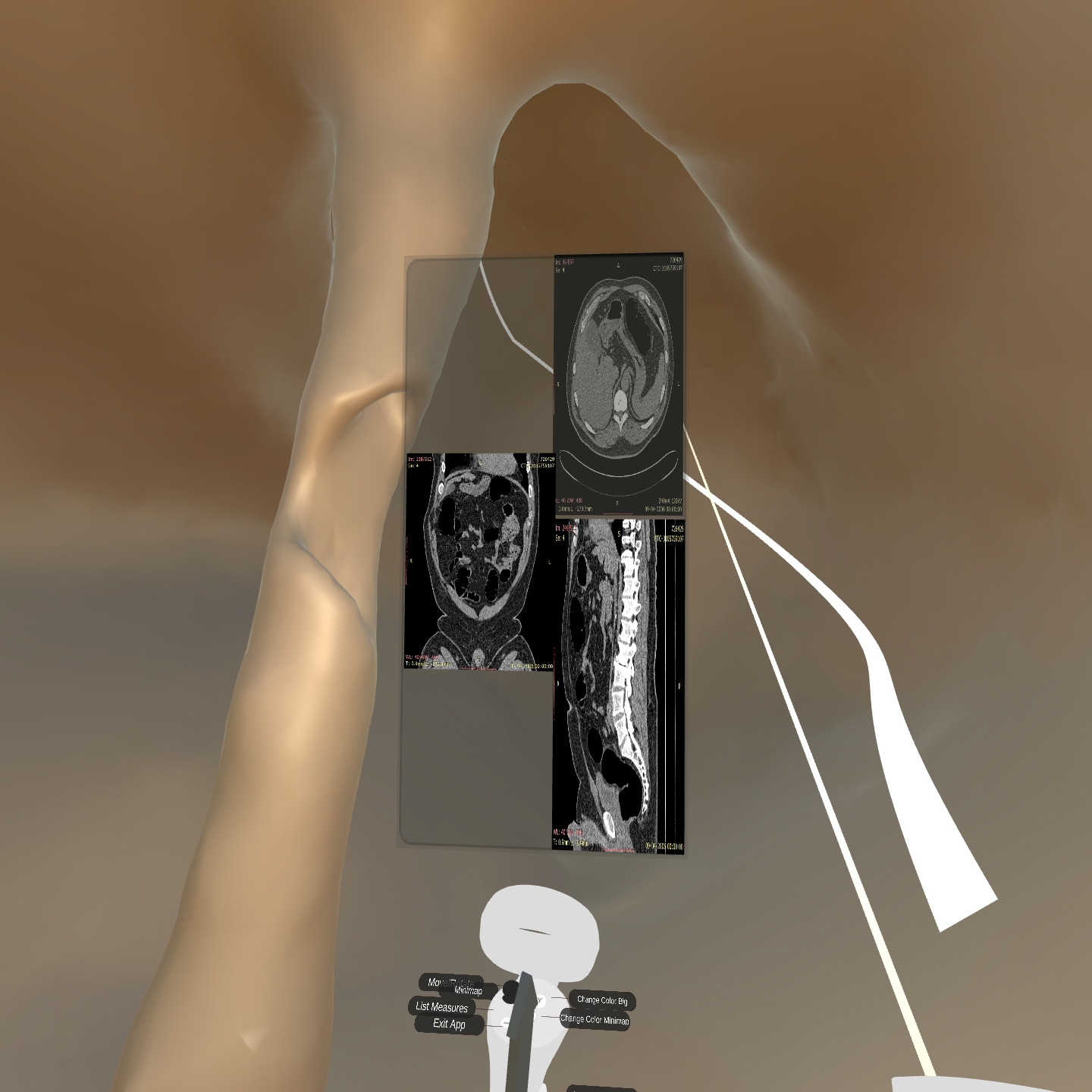}
\caption{Slice View}
\label{fig_slice}
\end{figure}

A final and crucial feature is to allow the analysis of the abnormality location on the CT scan. Instead of "splitting the screen", which is usual in desktop apps, to show three canonical planes, we propose to use the 3D space for that. In our current implementation, CT images can be seen on a floating panel upon pointing to a location (see fig.~\ref{fig_slice}). Moving the pointing ray causes the images to update for the new location. However, this is a critical feature that we wish to investigate experimentally. We will consider other approaches, such as registering the images with the model and constructing arbitrary planes interactively. 

\section{Final Comments}

The current development was exposed to experts for informal feedback. Several redesigns occurred in this iterative process. The next step is to detail a protocol for formal experimentation. The main goal is to study coverage, accuracy, and efficiency among conditions. We hypothesize that the immersive interface allows for a reduction in the time consumption for the exam while the false-positives and false-negatives rates remain similar to the desktop approach. 

A group of radiologists from more than one hospital will be recruited to perform an IVC on colons generated from public data with ground truth. They will perform the colonoscopy with only one run (rectum to cecum), two runs (rectum to cecum and cecum to rectum), and also on a conventional desktop VC with two runs (rectum to cecum and cecum to rectum) as it is the usual setup in clinical settings. In all tests, the area covered, the time consumed, and the characteristics of the anomalies discovered will be compared.

We recognize that a between-groups protocol is more suitable for this analysis to avoid memorization among conditions. Still, some of the variables will be tested within groups due to the restricted number of professional radiologists with experience in VC. Initially, we plan to expose the participants to all conditions but with different colons, randomizing the order of colons and interfaces.

The system will be installed on Pico Neo 3 Pro Eye HMD. Every action will be recorded for pattern analysis, and their reactions will be taken into account for discussion over the quantitative results. Action patterns and exam outcomes will also be crossed with subjective comfort and cybersickness.

\acknowledgments{
This work is partly supported by CNPq-Brazil project 311251/2020-0, by \textit{Fundação para a Ciência e Tecnologia} (Portuguese Foundation for Science and Technology) through grants 2022.09212.PTDC (XAVIER), UIDB/50021/2020, and Carnegie Mellon Portugal grant SFRH/BD/151465/2021 under the auspices of the UNESCO
Chair on AI\&XR.
}

\bibliographystyle{abbrv-doi}

\bibliography{template}
\end{document}